\documentclass[12pt,preprint]{aastex}

\usepackage{subfigure}
\usepackage{graphicx,times}
\usepackage{natbib}
\usepackage{amssymb,amsmath}
\usepackage[colorlinks,linkcolor=blue,citecolor=blue,urlcolor=blue]{hyperref}

\shorttitle{Meridional flow in the polar caps of the Sun}
\shortauthors{Yang et al.}

\begin{document}

\title{Meridional flow in the solar polar caps revealed by magnetic field observation and simulation}

\author{Shuhong Yang\altaffilmark{1,2,7}, Jie Jiang\altaffilmark{3}, Zifan Wang\altaffilmark{1,2,7}, Yijun Hou\altaffilmark{1,2,7}, Chunlan Jin\altaffilmark{1,2,7}, Qiao Song\altaffilmark{4,6}, Yukun Luo\altaffilmark{3}, Ting Li\altaffilmark{1,2,7}, Jun Zhang\altaffilmark{5}, Yuzong Zhang\altaffilmark{1,2,7}, Guiping Zhou\altaffilmark{1,2,7}, Yuanyong Deng\altaffilmark{1,2,7}, and Jingxiu Wang\altaffilmark{1,2,7}}

\altaffiltext{1}{National Astronomical Observatories, Chinese Academy of Sciences, Beijing 100101, China; \textcolor{blue}{shuhongyang@nao.cas.cn}}

\altaffiltext{2}{School of Astronomy and Space Science, University of Chinese Academy of Sciences, Beijing 100049, China}

\altaffiltext{3}{School of Space and Environment, Beihang University, Beijing 102206, China; \textcolor{blue}{jiejiang@buaa.edu.cn}}

\altaffiltext{4}{Key Laboratory of Space Weather, National Satellite Meteorological Center (National Center for Space Weather), China Meteorological Administration, Beijing 100081, China}

\altaffiltext{5}{School of Physics and Materials Science, Anhui University, Hefei 230601, China}

\altaffiltext{6}{Innovation Center for FengYun Meteorological Satellite (FYSIC), Beijing 100081, China}

\altaffiltext{7}{State Key Laboratory of Solar Activity and Space Weather, National Space Science Center, Chinese Academy of Sciences, Beijing 100190, China}

\begin{abstract}
As a large-scale motion on the Sun, the meridional flow plays an important role in determining magnetic structure and strength and solar cycle. However, the meridional flow near the solar poles is still unclear. The Hinode observations show that the magnetic flux density in polar caps decreases from the lower latitudes to the poles. Using a surface flux transport model, we simulate the global radial magnetic field to explore the physical process leading to the observed polar magnetic distribution pattern. For the first time, the high-resolution observations of the polar magnetic fields observed by Hinode are used to directly constrain the simulation. Our simulation reproduces the observed properties of the polar magnetic fields, suggesting the existence of a counter-cell meridional flow in the solar polar caps with a maximum amplitude of about 3 m s$^{-1}$.
\end{abstract}


\section{Introduction}

At the surface of the Sun, there are many kinds of flows such as granulation, supergranulation, differential rotation, and meridional flow.
The meridional flow is a large-scale motion in latitude on the Sun \citep{2022FrASS...907290H}.
Because the meridional flow is very weak, it is difficult to measure \citep{1996ApJ...460.1027H}. There are some methods to determine the meridional flow, e.g., direct Doppler-shift measurement, tracer tracking, and helioseismology analysis. In the early studies, a poleward meridional flow is suggested with direct Doppler observations \citep{1979SoPh...63....3D,1982SoPh...80..361L,1988SoPh..117..291U,1996ApJ...460.1027H}.
By following the small magnetic features that populate the solar surface, the poleward meridional flow is found at all latitudes up to 75$\degr$, and its speed is of order 10 m s$^{-1}$ \citep{1993SoPh..147..207K,2010Sci...327.1350H}.
The meridional flow at the solar surface and in the shallow interior can be detected with helioseismic methods \citep{1997Natur.390...52G,2004ApJ...603..776Z,2013ApJ...774L..29Z}.
Using the time-distance helioseismology, \cite{2020Sci...368.1469G} found that the time-averaged meridional flow is a single cell in each hemisphere, and the flow is equatorward at the base of the convection zone and poleward at the solar surface. However, the meridional flow near the solar poles is still unclear.

The meridional flow is deemed to play an important role in determining the magnetic structure and strength in polar regions \citep{2002ApJ...577L..53W,2009ApJ...693L..96J}. In the solar polar regions, there only exist small-scale magnetic structures, which are not easy to accurately measure in the limb observations. The pioneering polar magnetic field observations are only the measurement of the line-of-sight component of the magnetic field. The magnetic fields in the polar regions are assumed radial, and the magnetic strength is found to increase poleward and peaks at the poles at the epoch of solar cycle minimum \citep{1989ApJ...347..529W,2009ApJ...699..871P}. The vector magnetic field in the polar region was first systematically measured in 1997, and the net magnetic flux was found to approach roughly the interplanetary magnetic field \citep{1999ScChA..42.1096D}.
With the polar vector magnetic field data from the Hinode \citep{2007SoPh..243....3K} satellite, it is revealed that the vertical magnetic fields in many magnetic islands are stronger than one kilogauss within 70--90$\degr$ latitude \citep{2008ApJ...688.1374T}.
Then, one question is raised: which kind of meridional flow could result in the polar magnetic distribution pattern as observed by Hinode?

Surface flux transport (SFT) models have been remarkably successful in reproducing the magnetic field patterns at the solar surface \citep{2020ApJ...904...62W,2023SSRv..219...31Y}.
In these models, meridional flow, differential rotation, and supergranular diffusion are included \citep{1989ApJ...347..529W,2002ApJ...577.1006S,2004A&A...426.1075B}.
In the present study, we use an SFT model \citep{1989Sci...245..712W,2014ApJ...791....5J} to simulate the global radial magnetic field to explore the physical process leading to the observed polar magnetic field properties. For the first time, Hinode's high-resolution observations of the vector magnetic fields in polar caps are used to directly constrain the SFT simulation.

\section{Observed magnetic fields}

The Spectro-polarimeter (SP; \citealt{2013SoPh..283..579L}) of the Solar Optical Telescope \citep{2008SoPh..249..167T} on board Hinode provides the high spatial resolution and polarimetric precision measurement of the photospheric vector magnetic field. In order to study the magnetic fields in polar caps, we use the Hinode/SP data released at ISEE, Nagoya University\footnote[1]{https://hinode.isee.nagoya-u.ac.jp/sot\_polar\_field/}.
The pixel size along the east--west scanning direction is about 0{\arcsec}.30 and that along the slit is about 0{\arcsec}.32.
The SP full stokes profiles of Fe~{\sc{i}} 630.15 nm and 630.25 nm were processed to retrieve the vector magnetic field through stokes inversion with the Milne-Eddington inversion code \citep{2007A&A...462.1137O}.
The 180$\degr$ ambiguity of the transverse magnetic field was removed with the method of \cite{2010ApJ...719..131I}. More details can be found in \cite{2012ApJ...753..157S}.
Considering the optimal viewing, the observations for the south and north polar caps were mainly in March and September, respectively.
The SP data used in this study were obtained from 2012 to 2021, and every year for each pole, a batch of about 10 magnetograms with an interval of about 3 days is adopted \citep[see Table 1 of][]{2024RAA....24g5015Y}.

\begin{figure}
\centering
{\subfigure{\includegraphics[bb=46 377 547 458, clip,angle=0,width=\textwidth]{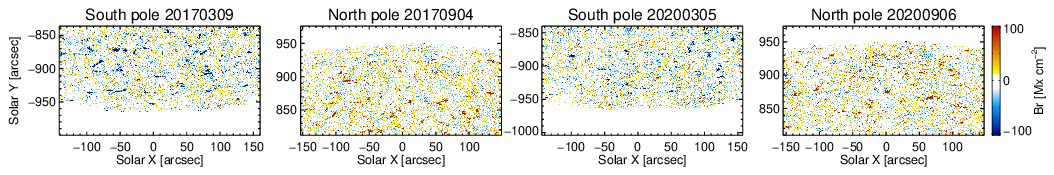}}\vspace{-0.08in}
\subfigure{\includegraphics[bb=46 355 547 495, clip,angle=0,width=\textwidth]{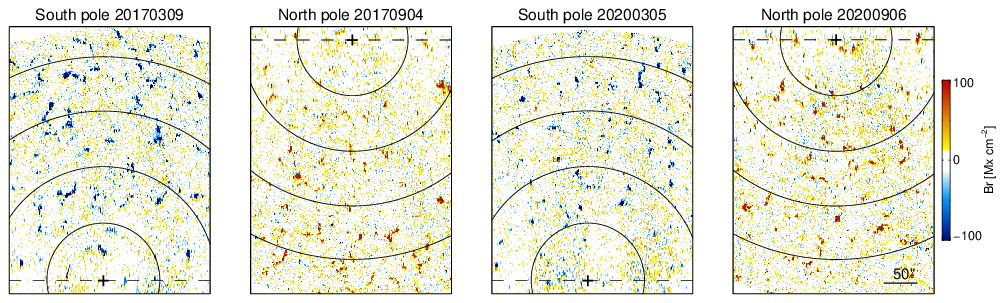}}}
\caption{Radial magnetic fields in the solar polar caps observed by Hinode/SP in 2017 and 2020. The upper panels and lower panels are radial magnetograms from the Hinode view and polar view, respectively. East is to the left and west is to
the right. The solid curves indicate the latitude separated by 5$\degr$, and the plus signs mark the poles. The dashed lines indicate the boundary, beyond which the region is not considered.}
\label{fig_BrMap}
\end{figure}

Figure \ref{fig_BrMap} displays the polar magnetic distribution observed in 2017 after the polarity reversal in solar poles and in 2020 around the solar cycle minimum of sunspot number. The upper panels show the radial magnetic fields in the polar caps from the view of Hinode. Both in 2017 and 2020, the magnetic fields in the south polar cap were dominated by the negative polarity, and the large concentrations with the positive polarity were quite rare. On the contrary, the dominant magnetic polarity in the north polar cap was positive. In March and September every year, the south pole and north pole are tilted toward us with an angle of about 7$\degr$. Therefore, the Hinode polar magnetograms cover the regions from about $\pm$67$\degr$ to the poles and beyond. To correctly display the size and spatial distribution of the polar magnetic concentrations, the observed magnetograms are mapped from the Hinode view to the polar view, i.e., viewing from above the south or the north pole. The dashed lines indicated the boundary, beyond which the polar region is not considered. It seems that the concentrations with strong field strength near the poles are somewhat less than those at the lower latitudes.

\begin{figure}
\centering
\includegraphics[bb=80 418 297 573,clip,angle=0,width=0.7\textwidth]{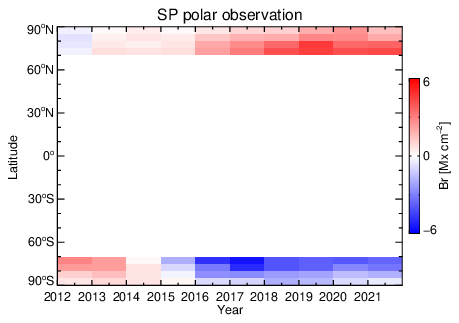}
\caption{Magnetic supersynoptic map created from Hinode/SP radial magnetograms.}
\label{fig_SP_Butterfly}
\end{figure}

In order to quantitatively study the magnetic fields at different latitudes, for the north and south polar caps, the averaged radial magnetic flux density within every 5$\degr$ (see the lower panels in Fig. \ref{fig_BrMap}) between $\pm$70$\degr$ and $\pm$90$\degr$ latitudes (excluding that beyond the poles) is calculated as $\sum_{ij}{Br_{ij}S_{ij}}/\sum_{ij}{S_{ij}}$, where $Br_{ij}$ is the radial flux density and $S_{ij}$ is the corrected real area in every pixel $i$ within each 5$\degr$ latitude range in each frame $j$ of the corresponding batch of $\sim$10 magnetograms every year. The averaged radial flux densities in 2017 and 2020 calculated from the Hinode/SP data are plotted with the color symbols in Fig. \ref{fig_BrSimulation}a and Fig. \ref{fig_BrSimulation}b, respectively. It is clear that both in the south and north polar caps, the magnetic flux density decreases from $\pm$70$\degr$ towards the poles.
We examine the ten-year SP observations and find that in each polar cap, during most of the time of the solar cycle except for the period around the polarity reversal, the higher the latitude, the weaker the radial magnetic flux density in general (as shown in Fig. \ref{fig_SP_Butterfly}).

We also adopt the radial magnetic field synoptic maps from the Helioseismic and Magnetic Imager (HMI; \citealt{2012SoPh..275..207S,2012SoPh..275..229S}) aboard the Solar Dynamics Observatory (SDO; \citealt{2012SoPh..275....3P}), available from the Joint Science Operations Center\footnote[2]{http://jsoc.stanford.edu/}. In the synoptic maps, the polar fields are temporally and spatially interpolated \citep{2011SoPh..270....9S,2018arXiv180104265S}.
The SDO/HMI magnetic flux density is multiplied by 1.18 to make it comparable with the Hinode/SP data \citep{2019RAA....19...69J}. The radial flux densities of every 5$\degr$ latitude range between $-$70$\degr$ and $+$70$\degr$ obtained from the SDO/HMI observations are also plotted in Fig. \ref{fig_BrSimulation}a-b, as represented with the black symbols. It shows that the maximum flux density is at the latitudes of about $\pm$70$\degr$.

\section{Simulated magnetic fields}

In this study, the SFT model uses the emergence of active regions (ARs) as sources of magnetic flux to simulate the evolution of the polar magnetic field.
The SFT model describes the evolution of radial magnetic field $B\left(\theta,\phi,t\right)$ at the solar photospheric surface $r=R_{\odot}$. The magnetic field is introduced onto the surface by a source term $S\left ( \theta ,\phi ,t \right )$, usually in the form of ARs, and then transported by differential rotation $\Omega \left ( \theta  \right )$, meridional flow $v\left ( \theta  \right )$, and diffused by supergranular diffusion $\eta$. Coordinates $\theta$ and $\phi$ are colatitude and longitude, respectively. The equation of the SFT model is the vertical component of the magnetic induction equation, which is as follows,
\begin{eqnarray}\label{eqSFT}
\begin{aligned}
  \frac{\partial B}{\partial t}=-\Omega \left ( \theta  \right )\frac{\partial B}{\partial \phi }-\frac{1}{R_{\odot }\sin \theta }\frac{\partial }{\partial \theta }\left [v\left ( \theta  \right ) B\sin \left ( \theta  \right ) \right ]+\\
  \frac{\eta }{R{_{\odot }}^{2}}\left [\frac{1}{\sin \theta } \frac{\partial }{\partial \theta }\left ( \sin \theta \frac{\partial B}{\partial \theta } \right ) +\frac{1}{\sin ^{2}\theta }\frac{\partial ^{2}B}{\partial \phi^{2}}\right ]+\\
  S\left ( \theta ,\phi ,t \right ).
\end{aligned}
\end{eqnarray}

For the surface differential rotation, we use the profile $\Omega(\theta)=13.38-2.30\cos^2\theta-1.62\cos^4\theta$, $\theta \in [0,\pi]$ in units of degree day$^{-1}$ \citep{1983ApJ...270..288S}. The supergranular diffusion $\eta$ is adopted as 500 km$^2$ s$^{-1}$ based on the constraint of \cite{2020ApJ...904...62W}. The diffusion value is within the range listed in Table 6.2 of \cite{Schrijver2000}. For meridional flow, we adopt a double-cell profile with a counter-cell at the polar cap \citep{2009ApJ...693L..96J}, which is as follows.
\begin{eqnarray}\label{eqDBcells}
v(\lambda) =
    \begin{cases}
        v_{0}\sin( \pi\lambda/\lambda_{0} ) & | \lambda | \leq \lambda_{0}\\
        v_{1}\sin( \pi( 90-| \lambda | )/( 90-| \lambda_{0} | ) ) & \lambda < -\lambda_{0} \\
        -v_{1}\sin( \pi( 90-| \lambda | )/( 90-| \lambda_{0} | ) ) & \lambda > \lambda_{0},\\
    \end{cases}
\end{eqnarray}
where $\lambda$ is latitude in degrees, and $\lambda_{0}$ is the starting latitude of the counter-cell. As shown, both the main-cell and the counter-cell are sinusoidal functions with maximum values of $v_{0}$ and $v_{1}$, respectively. We use $\lambda_{0}=70\degr$, $v_{0}=11$ m s$^{-1}$ (\citealt{1998ApJ...501..866V}), and a set of $v_{1}$ values for comparison. This profile is consistent with the two-term function given by \cite{Snodgrass1984} and \cite{1993SoPh..147..207K} for latitudes lower than $\pm$30$^\circ$. The divergence of the meridional flow at the equator represents a principal parameter influencing the total polar flux generated over the course of a cycle \citep{Petrovay2020, Jiang2023}.

The initial condition of the magnetic field is the radial magnetic field synoptic map of Carrington Rotation (CR) 2097 from HMI, resized to the simulation resolution. The magnetic source term consists of ARs during CRs 2097-2236, covering most of solar cycle 24. The ARs are identified and extracted from HMI synoptic maps with the AR identification method \citep{2010ApJ...723.1006Z,2020ApJ...904...62W}, and then smoothed and resized to the simulation resolution. They are then introduced into the simulation on the day it crosses central meridian on the Sun. The surface field is simulated from CR 2097 to CR 2236.
In our simulation of the global radial magnetic field, the initial synoptic map of CR 2097 began on 19 May 2010.
We numerically solve the equation with an SFT code based on \cite{2004A&A...426.1075B}. The code has 360$\times$180 spatial resolution and a time interval of one day. The spatial component is processed by spherical harmonics decomposition to up to $l = 63$, which roughly corresponds to the resolution of supergranulation. The temporal component is solved with a 4-th order Runge-Kutta method.

We use the coefficient of determination ($r^{2}$) and root mean squared error ($RMSE$) to compare the SFT simulation results of different counter-cell flow ($v_{1}$) value cases to the polar field observations of Hinode. For a series of observed values $y_{i}$ and simulated values $\hat{y_{i}}$, the values are $r^{2}=1-\sum_{i}{\left(y_{i}-\hat{y_{i}}\right)^{2}}/\sum_{i}{\left(y_{i}-\left\langle y_{i}\right\rangle\right)^{2}}$, and $RMSE=\sqrt{\left\langle\left(y_{i}-\hat{y_{i}}\right)^{2}\right\rangle}$, where the $< >$ brackets indicate mean values. An $r^{2}$ closer to 1 and a smaller $RMSE$ indicate that the model describes the observations better. We use Hinode polar magnetic field results from year 2017 to year 2021, and use the 4 latitude intervals on both hemispheres we consider, so there are 40 data points for each $v_{1}$ value case. The calculated $r^{2}$ and $RMSE$ are shown in Table \ref{tab1}.
When the $v_{1}$ values are set to 2.5 m s$^{-1}$, 3.0 m s$^{-1}$, and 3.5 m s$^{-1}$, the calculated $r^{2}$ values are 0.944970, 0.949609, and 0.948761, and $RMSE$ values are 0.730036 Mx cm$^{-2}$, 0.698592 Mx cm$^{-2}$, 0.704442 Mx cm$^{-2}$, respectively.
It means that, when assuming a counter-cell meridional flow from the pole to $\pm$70$\degr$ latitude with the maximum amplitude of 3 m s$^{-1}$, the simulation fits the observation well. The simulated flux densities as the function of latitude are displayed with the solid curves in Fig. \ref{fig_BrSimulation}a, b.

\begin{figure}
\centering
\includegraphics[bb=80 265 490 573,clip,angle=0,width=0.9\textwidth]{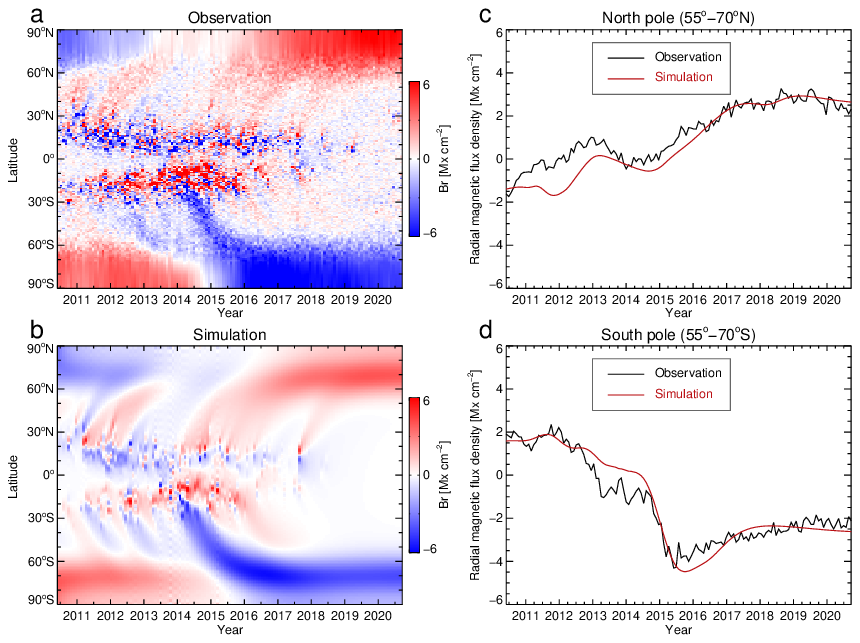}
\caption{Evolution of observed and simulated magnetic fields.
(a) Observed magnetic butterfly diagram generated from longitudinally averaged HMI synoptic maps.
(b) Simulated magnetic butterfly diagram with the SFT model assuming a counter-cell meridional flow of 3 m s$^{-1}$.
(c-d) Comparison between the simulated flux density and the HMI observed flux density averaged over latitudes 55$\degr$ $\leq \lambda \leq$ 70$\degr$ (north) and $-$70$\degr$ $\leq \lambda \leq$ $-$55$\degr$ (south), respectively.}
\label{fig_HMIvsSimu}
\end{figure}

Furthermore, in order to illustrate the reliability of our simulation, we compare the observed and simulated magnetic flux densities.
The simulated magnetic field is derived with the assumption of a counter-cell meridional flow of 3 m s$^{-1}$.
In the magnetic butterfly diagrams (Fig. \ref{fig_HMIvsSimu}a, b), the simulated magnetic fields at the mid-latitude and low-latitude are consistent with the HMI data. Even in the high latitude range $\pm$(55--70)$\degr$, where HMI can still measure the magnetic field well, the simulated fields are comparable to the HMI observations. This can be revealed by the comparison between the simulated and observed flux densities averaged over latitudes $\pm$(55--70)$\degr$ (Fig. \ref{fig_HMIvsSimu}c, d).
During the cycle minima, the simulated polar flux densities demonstrate a high degree of consistency with the observed values. The discrepancy between the observed and simulated polar field is predominantly evident in the northern hemisphere during the 2011-2013 period.  The difference could caused by a small number of inadequately identified ARs within activity complexes \citep{2020ApJ...904...62W}. These ARs are characterized by large area, high latitude, and considerable tilt angle with the orientation that was incongruous with the prevailing magnetic field configuration \citep{Jiang2015}, which lead to the alternative red and blue strong poleward surges shown in Fig. \ref{fig_HMIvsSimu}a.
For the polar regions greater than 70 degree latitude, we compare the SP observed (Fig. \ref{fig_SP_Butterfly}) and SFT simulated (Fig. \ref{fig_HMIvsSimu}b) supersynoptic maps, and find that the magnetic fields therein are comparable.
It means that our simulation is reasonable and reliable.

\begin{figure}
\centering
\includegraphics[bb=82 291 495 545,clip,angle=0,width=0.9\textwidth]{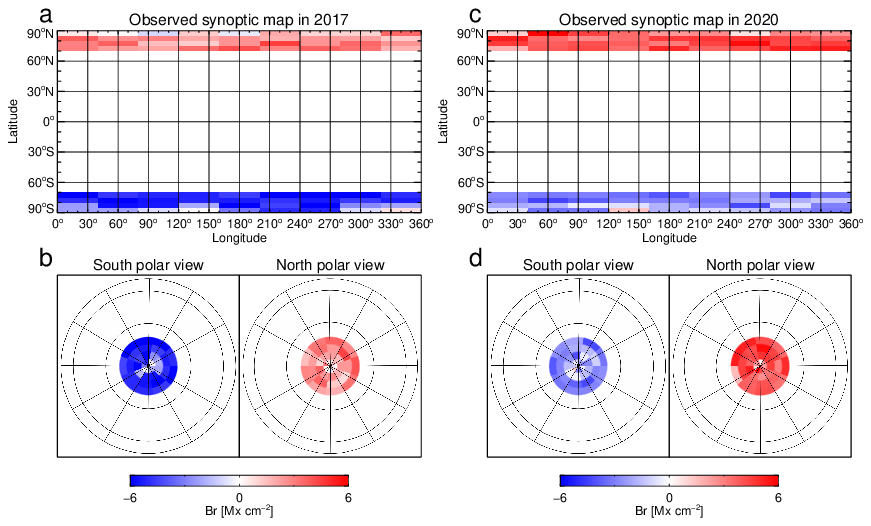}
\caption{Radial magnetic fields observed by Hinode/SP in March (the south polar cap) and September (the north polar cap). (a-b) Synoptic map observed in 2017 and corresponding re-mapped images viewing from above two poles. (c-d) Similar to (a-b) but for that observed in 2020. Both the longitude and latitude grids are separated by 30$^{\circ}$.}
\label{figure_SP_Synop}
\end{figure}

Consequently, using the simulation constrained by the SP observations, we can construct the synchronic maps of the radial magnetic field at any given time, especially for the weak magnetic field, which is often without good observation. The global radial magnetic fields, for example on 23 June 2017 and 11 August 2020, are mapped from the simulated synchronic images (Fig. \ref{fig_SynopSimu}a, c) on spherical surface viewing from above the solar poles (Fig. \ref{fig_SynopSimu}b, d), or from any free viewing angles (Fig. \ref{fig_BrSimulation}c, d, and the associated animation).
In order to gauge the quality of the simulated synchronic maps displayed in Fig. \ref{fig_SynopSimu}, we also construct the synoptic maps (see Fig. \ref{figure_SP_Synop}) from Hinode/SP magnetograms to compare with them. The simulated maps and observed maps both in 2017 and 2020 exhibit a consistency in the magnetic flux density decrease to the poles. In the south polar cap, the simulated magnetic field in 2017 is stronger than that in 2020. While in the north polar cap, the simulated magnetic field in 2017 is relatively weaker compared with that in 2020. These are all consistent with the Hinode/SP observations in 2017 and 2020.

\section{Conclusions and Discussion}

\begin{figure}
\centering
\includegraphics[bb=91 265 503 575, clip,angle=0,width=0.9\textwidth]{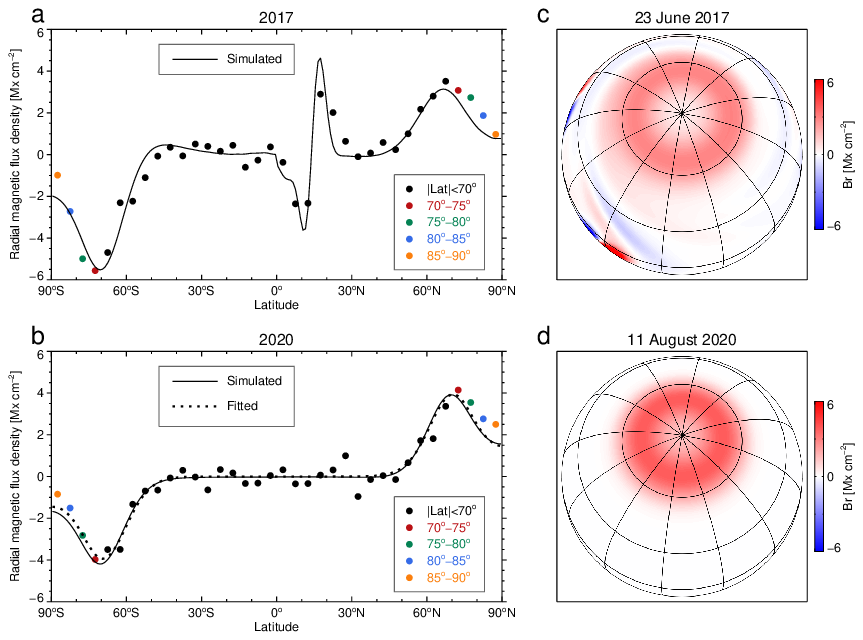}
\caption{Observation and simulation of radial magnetic flux density.
(a-b) Radial flux density as a function of latitude in 2017 and 2020, respectively. The color and black symbols represent the observations from Hinode/SP and SDO/HMI, respectively. The solid curves are the simulated results, and the dotted curve is the corresponding fitted result.
(c) Simulated synchronic image mapped on the spherical surface showing the radial magnetic flux density at the decline phase of solar cycle 24 on 23 June 2017.
(d) Similar to (c) but at the solar minimum on 11 August 2020.
The spherical images are viewed from above the 70$\degr$N latitude, and both the longitude and latitude grids are separated by 30$\degr$. (An animation of this figure is available. The animation shows the global radial magnetic fields mapped from the simulated synchronic images on the spherical surface. The dates are 23 June 2017 and 11 August 2020, respectively. Both the longitude and latitude grids are separated by 30$\degr$.)}
\label{fig_BrSimulation}
\end{figure}

\begin{figure}
\centering
\includegraphics[bb=82 291 495 545,clip,angle=0,width=0.9\textwidth]{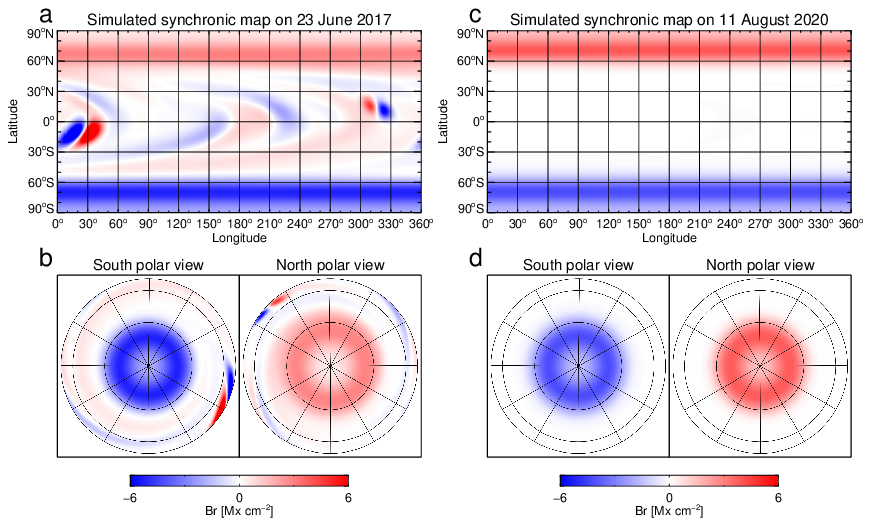}
\caption{Simulated radial magnetic fields with the assumption of a counter-cell meridional flow besides the main poleward flow.
(a-b) Simulated synchronic map on 23 June 2017 and the corresponding re-mapped images viewing from above two poles.
(c-d) Similar to (a-b) but on 11 August 2020.
Both the longitude and latitude grids are separated by 30$^{\circ}$. The northern polar views are also shown in Fig. \ref{fig_BrSimulation}.}
\label{fig_SynopSimu}
\end{figure}

For the meridional flows close to the poles, there is observational evidence that the flows are poleward during some parts of the solar cycle and equatorward during other epochs \citep{2010ApJ...725..658U,2022FrASS...907290H,2022RNAAS...6..181U}. The counter-cells have been observed during cycle minimum \citep{2010ApJ...725..658U}, which is also revealed to exist in our simulation (see Fig. \ref{fig_BrSimulation}).
For the simulated radial magnetic field (solid curve in Fig. \ref{fig_BrSimulation}b) at the solar minimum, it can be fitted (dotted curve in Fig. \ref{fig_BrSimulation}b) by the following formula:
\begin{eqnarray}\label{eqFit}
 B_{\rm r}(R_\odot,\lambda)=  \begin{cases} {\rm sgn} (\lambda) B_{\rm 0} {\rm exp} \{ -a_{\rm 0}[{\rm cos}(\pi \lambda/\lambda_{\rm 0}) +1] \} &  |\lambda| \leq \lambda_{\rm 0}\\
  {\rm sgn} (\lambda) B_{\rm 0} {\rm exp} \{ a_{\rm 1} [{\rm cos}(\pi (| \lambda|-\lambda_{\rm 0})/(90-\lambda_{\rm 0}))-1]\} & |\lambda| > \lambda_{\rm 0}, \end{cases}
 \end{eqnarray}
where $\lambda \in $ [-90, 90]$\degr$ is the latitude in degrees, $\lambda_{\rm 0}=70\degr$, ${\rm sgn}(\lambda)$ is the sign of the latitude, $B_{\rm 0}$ is the maximum of polar field strength, $a_{\rm 0} = v_{\rm 0} R_\odot \lambda_{\rm 0} / (\pi \eta)$, $a_{\rm 1} = v_{\rm 1} R_\odot (90 - \lambda_{\rm 0}) / (\pi \eta)$, $v_{\rm 0}$ is the poleward meridional flow velocity, and $v_{\rm 1}$ is the counter-cell meridional flow velocity. For $|\lambda| \leq \lambda_{\rm 0}$, the description is given by \cite{1998ApJ...501..866V}. While for $|\lambda| > \lambda_{\rm 0}$, the description is given for the first time in the present study.
Equation (\ref{eqFit}) is obtained by assuming that the diffusion of flux balances the meridional flow advection at the cycle minimum, similar to how \cite{1998ApJ...501..866V} obtained the $|\lambda| \leq \lambda_{\rm 0}$ part. Since it is based on the SFT processes, its form holds as long as the meridional flow follows Eq. (\ref{eqDBcells}). Equation (\ref{eqFit}) implies that the magnetic field at $|\lambda| > \lambda_{\rm 0}$ decreases monotonically as we increase the $v_{\rm 1}$ value to larger than 3.5 m s$^{-1}$ listed in Table \ref{tab1}. Therefore, it is obvious that $v_{\rm 1} =$ 3 m s$^{-1}$ is the best fit leading to a maximum $r^{2}$ and a minimum $RMSE$ value.

\begin{figure}
\centering
\includegraphics[bb=90 265 342 575,clip,angle=0,width=0.6\textwidth]{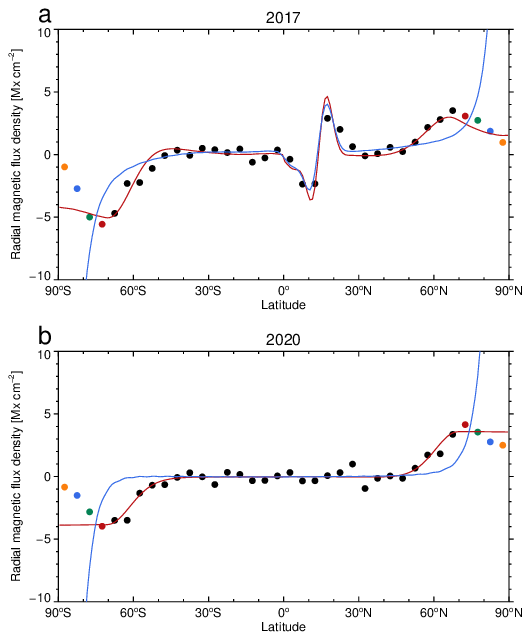}
\caption{Comparison between simulated and observed flux densities assuming no counter-cell flow.
(a) Observed radial magnetic flux densities (black and color symbols) in 2017 and simulated ones with the assumptions of a single-cell poleward surface flow stopping at $\lambda =$ 70$\degr$ (red curve) and a single-cell poleward surface flow from the equator to the pole (blue curve), respectively. (b) Similar to (a) but in 2020.}
\label{fig_BrSimulation_assumption}
\end{figure}

We also perform two simulations assuming no counter-cell flow: a single-cell poleward flow stopping at $\pm$70$\degr$ latitude, and a single-cell poleward flow stopping at $\pm$90$\degr$ latitude.
The profile of a single-cell poleward flow stopping at $\pm$70$\degr$ latitude is described as
\begin{eqnarray}\label{eqNoC}
v(\lambda) =
    \begin{cases}
        v_{0}\sin( \pi\lambda/\lambda_{0} ) & | \lambda | \leq \lambda_{0}\\
        0 & {\rm otherwise},\\
    \end{cases}
\end{eqnarray}
where $v_{0}=11$ m s$^{-1}$, $\lambda$ is latitude in degrees, and $\lambda_{0}=70\degr$ is the latitude above which the single-cell poleward flow vanishes.
The profile of a single-cell poleward flow extending to the pole is described as
\begin{eqnarray}\label{eqToPole}
v(\lambda) = v_{0}\sin( \pi\lambda/\lambda_{0} ),
\end{eqnarray}
where $v_{0}=11$ m s$^{-1}$, $\lambda$ is latitude in degrees, and $\lambda_{0}=90\degr$.
As shown in Fig. \ref{fig_BrSimulation_assumption}, if there is no meridional flow at latitudes higher than $\pm$70$\degr$, as the latitude increases, the polar magnetic flux density derived from the simulation (represented by the red curve in Fig. \ref{fig_BrSimulation_assumption}a) at the decline phase of solar cycle 24 will decrease more slightly than that from the observation. In particular, the simulated radial magnetic flux density at the solar minimum is uniform in the polar region above $\pm$70$\degr$ latitude (see the red curve in Fig. \ref{fig_BrSimulation_assumption}b).
The calculated $r^{2}$ and $RMSE$ are 0.769231, and 1.49497 Mx cm$^{-2}$, respectively (see Table \ref{tab1}).
If there exists a poleward meridional flow to the pole, the simulated magnetic flux density will increase as the latitude increases from $\pm$70$\degr$ to $\pm$90$\degr$ (see the blue curves in Fig. \ref{fig_BrSimulation_assumption}). The calculated $r^{2}$ and $RMSE$ are $-$145, and 37 Mx cm$^{-2}$, respectively. Both of these simulation cases are inconsistent with the SP observations.

\begin{figure}
\centering
\includegraphics[bb=77 320 495 530,width=0.9\textwidth]{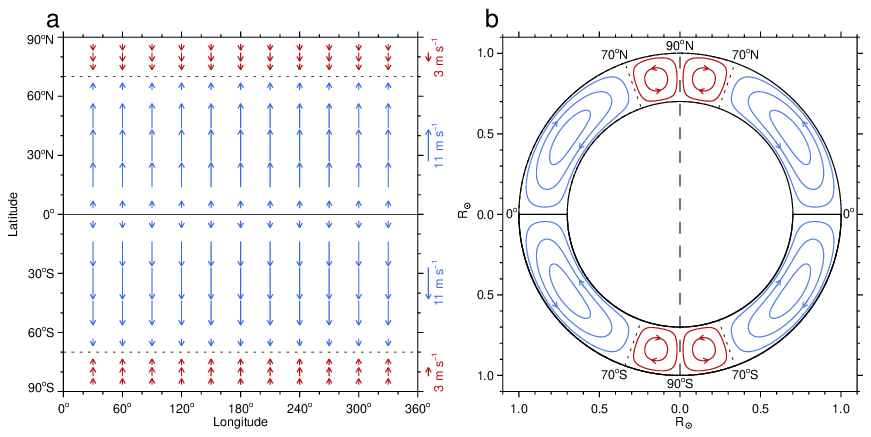}
\caption{Schematic cartoon showing the meridional flow pattern of the Sun.
(a) Velocity distribution of the meridional flows at the solar surface.
The blue and red arrows represent the poleward meridional flow and the equatorward meridional flow, respectively.
(b) Meridional flows as functions of latitude and depth assuming a single cell in radius. The blue and red closed loops represent the main-cell and the counter-cell, respectively.}
\label{fig_Cartoon}
\end{figure}

Based on the observation and simulation, we can deduce the meridional flow pattern of the Sun (see Fig. \ref{fig_Cartoon}).
Figure \ref{fig_Cartoon}a shows the meridional flow used in the SFT model and the deduced one for the poleward latitude range. They are constructed as functions of latitude averaged in longitude and time. Both the poleward meridional flow (indicated by the blue arrows) and the equatorward one (indicated by the red arrows) in each hemisphere are sinusoidal functions with maximum values of $v_{0}$ and $v_{1}$, respectively. They are shown as functions of longitude from pole to pole, implying that there is no variation with longitude.
The poleward meridional flow is located within $\pm$(0--70)$\degr$ latitude with the maximum amplitude of $v_{0} =$ 11 m s$^{-1}$ at $\pm$35$\degr$ latitude and the equatorward meridional flow is located within $\pm$(70--90)$\degr$ latitude with the maximum amplitude of $v_{1} =$ 3 m s$^{-1}$ at $\pm$80$\degr$ latitude.
Figure \ref{fig_Cartoon}b shows meridional flows as functions of latitude and depth with the assumption of a single cell in radius.
Mass conservation requires that the plasma should be carried around cells \citep{2011Natur.471...80N,2020Sci...368.1469G}. For the geometry of the meridional flow in the solar convection zone, previous studies have given different conclusions, i.e., either one cell or two cells in the radial direction \citep{2013ApJ...774L..29Z,2020Sci...368.1469G}.
The observational evidence here is inconclusive with regard to the number of cells in radius. If we assume that there is a single cell in the radial direction, then the flow pattern is deduced to consist correspondingly of two meridional circulation cells in each hemisphere, forming a double-cell flow pattern in the solar convection zone,
as shown in Fig. \ref{fig_Cartoon}b. The main-cell (represented by the blue closed loops) and counter-cell (represented by the red closed loops) meridional flows meet at the solar surface with the latitude of $\pm$70$\degr$ and return in the subsurface via equatorward and poleward flows, respectively. It should be pointed out that the density changes by several orders of magnitude in the solar convection zone. As a consequence, the amplitudes of meridional flows will be vastly different between the solar surface and the bottom of the convection zone.

The present study is based on the data observed in the plane of the ecliptic, and thus the polar magnetic field data have some errors in the limb observations \citep{2017SoPh..292...13P,2022ApJ...941..142P}. If there are more reliable measurements of the polar magnetic fields in the future, the current main result that there exists a counter-cell meridional flow in the polar caps with a maximum amplitude of about 3 m s$^{-1}$ may be changed.
The recently launched Solar Orbiter \citep{2020A&A...642A...1M}, and the planned missions, e.g., the Solaris \citep{2023arXiv230107647H}, and the Chinese Solar Polar-orbit Observatory \citep{2023TB..68..298D}, are designed to travel in large solar inclination angles and measure the polar regions unprecedentedly well. In future, the observations obtained from a polar vantage point will eventually help us to well understand the polar caps of the Sun.

\acknowledgments {We thank Prof. Yukio Katsukawa at NAOJ for the helpful discussion.
This research is supported by the National Key R\&D Programs of China (2022YFF0503800, 2022YFF0503000, 2019YFA0405000), the Strategic Priority Research Programs of the Chinese Academy of Sciences (XDB0560000, XDB41000000), the National Natural Science Foundations of China (12173005, 12273060, 12350004, 12273061, 12222306, 12073001), the Youth Innovation Promotion Association CAS, and Yunnan Academician Workstation of Wang Jingxiu (No. 202005AF150025).
The data are used courtesy of Hinode and SDO teams.
Hinode is a Japanese mission developed and launched by ISAS/JAXA, with NAOJ as domestic partner and NASA and STFC (UK) as international partners. It is operated by these agencies in co-operation with ESA and NSC (Norway). ISEE Database for Hinode SOT Polar Magnetic Field (doi: 10.34515/DATA\_HSC-00001) was developed by the Hinode Science Center, Institute for Space-Earth Environmental Research (ISEE), Nagoya University.
SDO is the first mission of NASA's Living With a Star Program.
}

\clearpage

\begin{table}
\begin{center}
\caption{Calculated $r^{2}$ and $RMSE$ \label{tab1}}
{\vskip 5mm}
 \begin{tabular}{ccc}
\tableline
$v_{1}$ (m s$^{-1}$) & $r^{2}$ & $RMSE$ (Mx cm$^{-2}$) \\
\tableline
2.5 & 0.944970 & 0.730036 \\
3.0 & 0.949609 & 0.698592 \\
3.5 & 0.948761 & 0.704442 \\
0 & 0.769231 & 1.49497 \\
To pole\tablenotemark{a} & -145 & 37 \\
\tableline
\end{tabular}
\tablenotetext{a}{A single-cell poleward flow extending to the pole.}
\end{center}
\end{table}

\clearpage

\clearpage

\clearpage

\clearpage

\clearpage

\clearpage

\clearpage

\clearpage

\clearpage

\end{document}